\documentstyle[12pt,epsf]{article}

\title{A scalar field modelling of the rotational curves of spiral galaxies}

\author{J.P.  Mbelek \\ Service d'Astrophysique, C.E.  Saclay \\ F-91191
Gif-sur-Yvette Cedex, France}

\begin{document} \maketitle \baselineskip=8mm

\begin{abstract} In a previous work \cite{mbeleka}, we have modelled the
rotation curves (RC) of spiral galaxies by including in the equation of motion
dynamical terms from an external real self-interacting scalar field, $\phi$,
minimally coupled to gravity and which respects the equivalence principle in the
absence of electromagnetic fields.  This model appears to have three free
parameters :  the turnover radius, $r_{0}$, the maximum rotational velocity,
$v_{max} = v(r_{0})$, plus a strictly positive integer, $n$.  Here, the coupling
of the $\phi$-field to other kinds of matter is emphasized at the expense of its
self-interaction.  This reformulation presents the very advantageous possibility
that the same potential may be used now for all galaxies.  New correlations are
established.  \end{abstract}

\section{Introduction} The equation of motion of a neutral test body reads in
the presence of the $\phi$-field \begin{equation} \label{eq motion}
\frac{du^{\mu}}{ds} \,+ \,{\Gamma}^{\mu}_{\alpha\beta} \,u^{\alpha} \,u^{\beta}
= \frac{d\phi}{ds} \,u^{\mu} \,- \,{\partial}^{\mu} \phi, \end{equation} where
the geodesic equation is recovered in the case of a non variable $\phi$-field.
In the weak fields and low velocity limit, assuming spherical symmetry and
circular orbits, the rotational velocity, $v$, at radius $r$ is then given by
\begin{equation} \label{angular momentum vs phi} \ln{(rv)} = \ln{J} + \phi,
\end{equation} where $J$ is a constant which would represent the angular
momentum per unit mass if the $\phi$-field were not present within the
galaxy\footnote{Here, we just deal with the tangential equation since we have
assumed circular orbits.  The radial equation would yield the radial velocity
but, it is found negligible with respect to the tangential velocity, out of the
bulge.}.  Assuming that the excitation of the $\phi$-field is very small
compared to its vacuum expectation value, ${\phi}_{vev}$, the equation of the
$\phi$-field reads in the first approximation \begin{equation} \label{eq phi}
{\partial}_{\mu}\,{\partial}^{\mu} \phi = - \,g \,\chi \,{\phi}_{vev} \,T,
\end{equation} where $T$ denotes the trace of the energy-momentum tensor of the
source of the $\phi$-field, $\chi = 8\pi G/c^{4}$ is the Einstein gravitational
constant and $g$ a universal dimensionless coupling constant.  Hence, one gets
for a static spherical matter distribution \begin{equation} \label{eq phi static
spherical} \frac{d^{2}\phi}{dr^{2}} \,\,+ \,\,\frac{2}{r} \,\frac{d\phi}{dr} = g
\,\frac{8\pi G}{c^{2}} \,{\phi}_{vev} \,\rho, \end{equation} where $\rho =
{\rho}_{bulge} \,+ {\rho}_{disk} \,+ \,{\rho}_{halo}$ denotes the mass density
of the matter fields other than the $\phi$-field itself.  Assuming in addition a
sufficiently thin disk (stellar plus gaseous disks) and a (quasi-isothermal)
spherical dark halo with mass density such that \begin{equation} \label{halo
density} {\rho}_{halo} \propto 1/r^{2 \,+ \,1/n}, \end{equation} the static
spherical solution of $\phi$ is found proportional to $r^{-1/n}$ (up to the
vacuum expectation value) within the galaxy out of the bulge.  Hence, the
rotational velocity reads \begin{equation} \label{velocity vs radius} v =
v_{max} \,G_{n}(r/r_{0}), \end{equation} where the functions $G_{n}$ are defined
as follows \begin{equation} \label{universal RC} G_{n}(x) = \frac{1}{x} \,\exp{[
\,n(1 - x^{-1/n}) \,]}.  \end{equation} For our purpose, $n \geq 2$.  Indeed,
this is one of the two necessary conditions for the $\phi$-field mimics a dark
matter mass profile ($\propto r^{1-1/n}$).

\section{Results} As can be seen in figure 1, the greater is the integer $n$,
the steepest is the curve $y = v/v_{max}$ versus $x = r/r_{0}$ for $x < 1$ and
the flatest it is for $x > 1$.  Hence, the steepest is a RC below the turnover
radius, the flatest it should be beyond.  Figure 2 shows some fits to individual
rotation curves (from the samples of Rubin {\it et al.}  \cite{rubina, rubinb},
van Albada {\it et al.}  \cite{van albada}, Lake and van Gorkom \cite{lake}).
This is achieved by using the least-squares fit to search the parameters $a$ and
$b = \ln{J}$ that yield the maximum square, $R^{2}$, of the correlation
coefficient for the relation $\ln{(rv)} = a r^{-1/n} + b$.  As yet, based on the
study of a hundred spirals, it is found that $a$ is always negative (this is the
other necessary condition for the $\phi$-field mimics a dark matter mass
profile) whereas $b$ is always positive.  In addition, one gets approximately
the following statistics :  $n = 5$ for $20$\% of the spirals, $3 \leq n \leq 6$
for almost a half of them and $3 \leq n \leq 12$ for $80$\% of them.  Figure 3
shows our new result :  There is a strong correlation between the coefficients
$a$ and $b$.  Though displayed here for only fifteen spirals, this remains true
for the whole sample of a hundred spirals we have studied hitherto.  Hence, if a
natural theoretical explanation is found in the future, only two independent
parameters will be needed to fit the RC with the proposed model one of which is
just an integer.

\section{Conclusion} It is possible that long range scalar fields external to
gravity play a significant role not only at the cosmological level but also at
the scale of galaxies or even the solar system \cite{mbelekb}.

\begin{figure} \centerline{\epsfxsize=12cm \epsfbox{mbelek1.epsf}}
\caption{Generic curves $y = G_{n} (x)$ for $n = 2, 4, 8, 16, 32$ and $64$.  As
one can see, the greater is $n$ the steepest is the curve below $x = 0.5$ and
the flatest it is beyond.}  \end{figure}

\begin{figure} \centerline{\epsfxsize=12cm \epsfbox{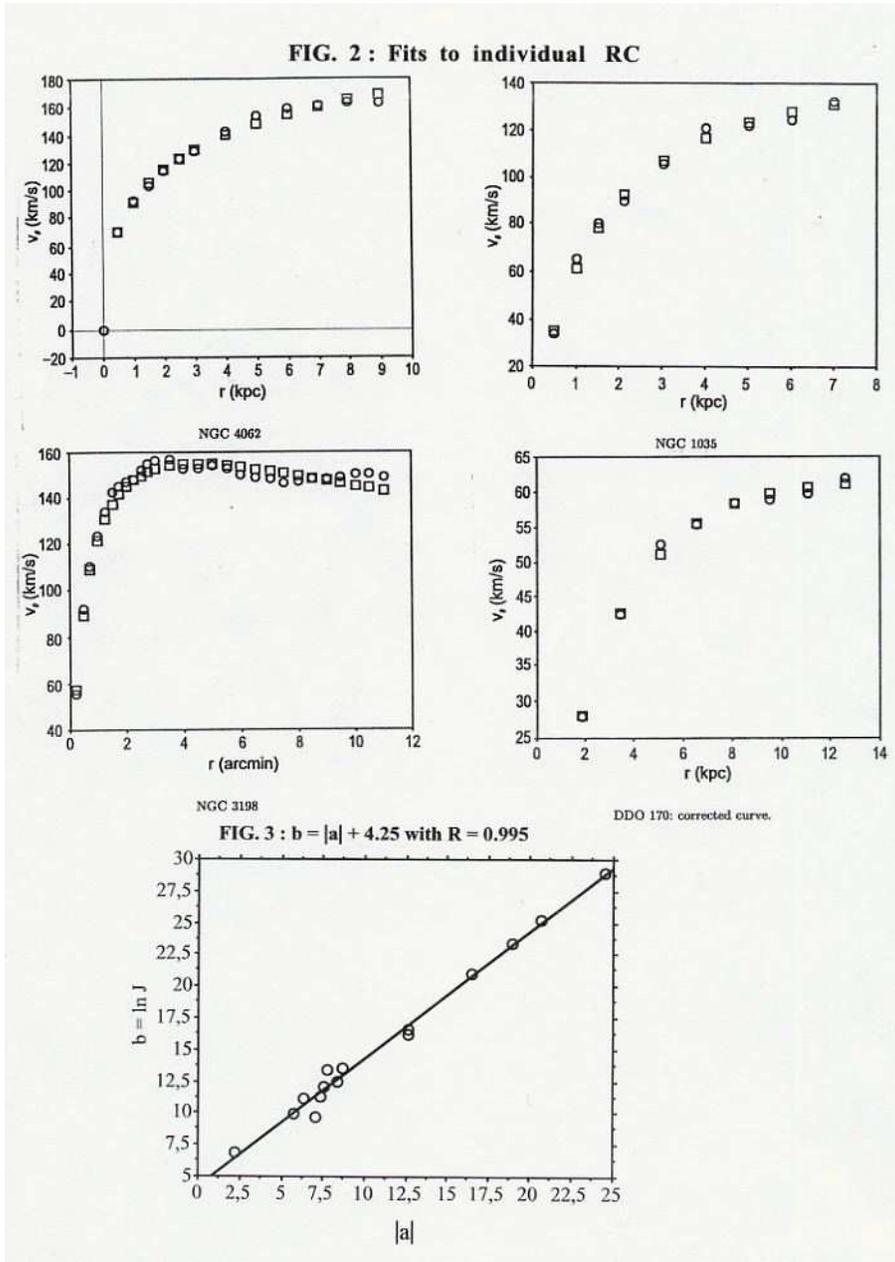}}
\caption{Rotation curve fits for NGC 4062~[2], NGC 1035~[3], NGC 3198~[4] and
DDO 170~[5].  Below, in figure 3 one can see the strong linear correlation
between $\mid a \mid$ and $b$.}  \end{figure}


\begin{thebibliography}{} \bibitem[1]{mbeleka} J.  P.  Mbelek, (1998) Acta
Cosmologica \textbf{XXIV-1}, 127 and gr-qc/0402084.

\bibitem[2]{rubina} Vera C.  Rubin, W.  K.  Jr.  Ford and N.  Thonnard, (1980)
ApJ \textbf{238}, 471.

\bibitem[3]{rubinb} Vera C.  Rubin, D.  Burstein, W.  K.  Jr.  Ford and N.
Thonnard, (1985) ApJ \textbf{289}, 81.

\bibitem[4]{van albada} T.  S.  van Albada, J.  N.  Bahcall, K.  Begeman and R.
Sancisi, (1985) ApJ \textbf{295}, 305.

\bibitem[5]{lake} G.  L.  Lake and J.  H.  van Gorkom, (1990) AJ \textbf{99},
547.

\bibitem[6]{mbelekb} J.  P.  Mbelek and M.  Lachi\`eze-Rey, (1999) SISSA
preprint gr-qc/9910105.

\end{thebibliography}
\end{document}